\documentclass[prl,showpacs,floatfix,aps]{revtex4}
\usepackage{graphicx}
\begin{document}
\title{Elastic constants of borocarbides. New approach to acoustic
measurement technique}

\author{E.\ A.\ Masalitin, V.\ D.\ Fil,\thanks{E-mail: fil@ilt.kharkov.ua}
K.\ R.\ Zhekov, A.\ N.\ Zholobenko, and T.\ V.\ Ignatova}

\address{B. Verkin Institute for Low Temperature Physics and
Engineering, National Academy of Sciences of Ukraine, pr.\ Lenina 47,
61103 Kharkov, Ukraine}

\author{Sung-Ik Lee}

\address{Department of Physics, Pohang University of Science and
Technology, Pohang 790-784, Korea}

\date{Published \ Low\ Temp.\ Phys. {\bf 29}, No.\ 1 (2003)}

\begin{abstract}

A new version of the phase method of determining the sound velocity is
55proposed and implemented. It utilizes the ``Nonius'' measurement
technique and can give acceptable accuracy ($\leq 1\%)$ in
samples of submillimeter size. Measurements of the sound velocity are
made in single-crystal samples of the borocarbides
RNi$_{2}$B$_{2}$C (${\rm R = Y,Lu,Ho}$). The elastic constants and
the Debye temperature are
calculated.

\end{abstract}
\pagenumbering{arabic}\thispagestyle{myheadings}

\maketitle

\section*{1. INTRODUCTION}

An important problem of physical acoustics is to obtain reliable data
on the elastic constants of newly synthesized compounds. These data,
while being of independent interest, also serve as tests for
theoretical calculations of band structures, force constants, and
phonon spectra. As a rule, newly synthesized materials come
either in the form of products of solid-phase synthesis (i.e., more or
`less porous ceramics) or in the form of fine single crystals. Objects
of the first group are characterized by appreciable scattering of
elastic vibrations, making it practically impossible to use some
version of a resonance or quasiresonance (of the long-pulse type)
method to determine the absolute values of the sound velocity in them.
Single crystals most often are of millimeter or submillimeter size;
besides, in layered crystals the characteristic size in the direction
perpendicular to the layers is often 100--200 $\mu$m or even less. To
determine the elastic constants of such objects the method of
ultrasonic resonance spectroscopy\cite{1} was developed, which consists
in measurement of the spectrum of resonance frequencies of a sample and
subsequent solution of the inverse problem of recovering all the
components of the tensor of elastic constants. The technique
is inherently a resonance method, i.e., it applies only to objects
with small scattering (damping), a condition which is not always
possible to satisfy even in small single crystals, e.g., near points of
phase transitions. In addition, it can be implemented only in samples
having a definite simple geometric shape (rectangular parallelepiped).
The lucidity of this method is compromised by the complexity of
the mathematical processing, making it hard to spot possible errors.

We have implemented a new version of the phase method of measuring
sound velocities; it is applicable both to ceramic samples with strong
scattering and to single crystals of submillimeter size. Utilizing a
kind of ``Nonius'' measurement procedure, the method permits one to
achieve acceptable accuracy (as a rule, better than 1\%) in both cases.
It has been used to measure the sound velocity in MgB$_{2}$
polycrystals\cite{2} and in VSe$_{2}$ layered single crystals.\cite{3}
Furthermore, being completely independent of the nature of the signals
to be analyzed, the instrumental implementation of the method enables
one to study the variation of the amplitude and phase of any pulsed
high-frequency signals. In particular, it has been used to measure the
characteristics of an electric field accompanying a longitudinal sound
wave in a metal.\cite{4}

Section 2 of this paper is devoted to a description of the basic
principles of implementation of the ``Nonius'' method of phase
measurements of sound velocity. In Sec.\ 3 we present the results of
measurements of the elastic constants in single crystals of the
borocarbides RNi$_{2}$B$_{2}$C (${\rm R = Y,Lu,Ho})$.

\section*{2. ``NONIUS'' METHOD OF SOUND VELOCITY MEASUREMENT.
PRINCIPLES AND INSTRUMENTAL IMPLEMENTATION}

\begin{figure}
\label{f1}
\includegraphics[width=15cm,angle=0]{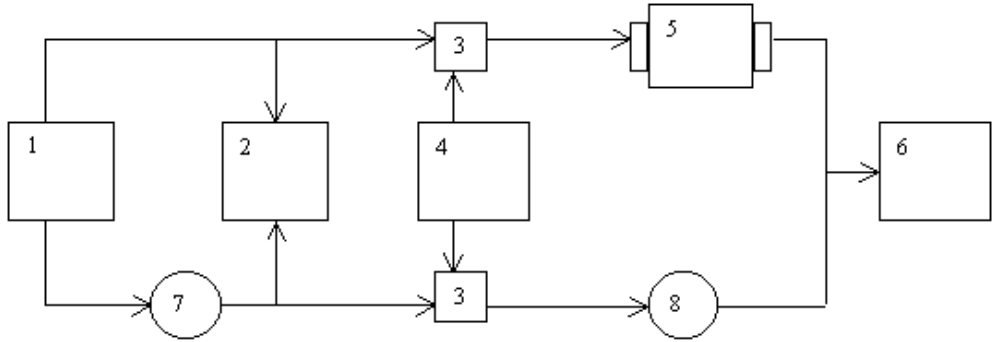}
\caption[]{Block diagram of the instrument: 1 --- frequency
synthesizer, 2 --- phase meter, 3 --- switches, 4 --- pulse-code modulation
unit, 5 --- sample with piezotransducers, 6 --- receiver, 7 ---
electronically tunable phase shifter, 8 --- smoothly adjustable
attenuator. }
\end{figure}

A block diagram of a device implementing the technique is presented in
Fig.\ \ref{f1}.

It is essentially a standard compensation or bridge
circuit, depending on the algorithm used for process the pulsed signals,
which is set by the pulse code modulation unit. In the bridge mode the signal
that passed the sample channel is summarized with the equal in absolute value
antiphase comparison one. The amplitude and phase
of the latter are regulated by the receiver, which functions as a null
device. The unbalance signal is separated into amplitude and phase
components by high-frequency synchronous detectors.\cite{5} In the
compensation mode the receiver, with the aid of sampling--storage
devices, matches the amplitudes of the signals arriving at its input
at different times.
In this case the unbalance signals with respect to amplitude
and phase are produced through a special code modulation of the
pulse trains of the signals in the two channels. In any variant the data
input to the computer are the readings of an attenuator
(amplitude of the comparison signal) and phase meter (phase
difference of the signal to be analyzed and the comparison signal).

Two original developments employed in the implementation of this
standard scheme have substantially expanded its operational
capabilities: an electronically controlled (linear) phase shifter
with a practically unlimited tuning range, and a new data processing
algorithm, which maintains a phase shift of 120$^{\circ}$ (or
240$^{\circ})$ between the signals being analyzed. The advantages of
the new phase shifter are quite obvious. In particular, in relative
measurements this phase shifter provides a practically unlimited
dynamic range while maintaining an extremely high accuracy of
measurement, which is actually determined by the resolution of the
phase meter (at a signal-to-noise ratio $ \geq 5$). Let us discuss
the second development in somewhat more detail. In the bridge mode the
working algorithm of the circuit consists in maintaining a null signal
at the input of the receiver upon changes in the sound velocity and
damping in the sample. In inhomogeneous (e.g., polycrystalline) samples,
internal re-reflections and mutual conversion of different modes at
inhomogeneities lead to nonconstancy of the phase of the signal over
the duration of the rf pulse envelope. The same situation is observed
in short single crystals due to the superposition of secondary
reflections. In this case the length of the time interval during
which the sum of the two signals has zero amplitude turns out
to be short ($ \leq 10^{-7}$ s). For analysis of such narrow
features the receiving system should have a rather wide passband
and not allow any overshoots under reproducing steep signal fronts.

For the 120$^{\circ}$ algorithm the sum of two signals of identical
amplitude (their equality is maintained by an independent channel) is
equal to the amplitude of each of the signals (equilateral triangle).
In this case at the time of sampling--storage there are no sharp
amplitude drops at the input of the receiver; this substantially
improves the working of the system as a whole. A distinct advantage of
the 120$^{\circ}$ algorithm is that it is unnecessary to have frequency
(phase) modulation of the master oscillator in order to obtain
unbalance signals of different polarity upon passage through the
compensation point, as one must have for self-balancing of the
circuit. Furthermore, the usual amplitude detection used in the
120$^{\circ}$ algorithm allows one to use as the signals of the two
channels any two reflections that have traveled different distances in
the sample.

\begin{figure}
\includegraphics[width=7cm,angle=0]{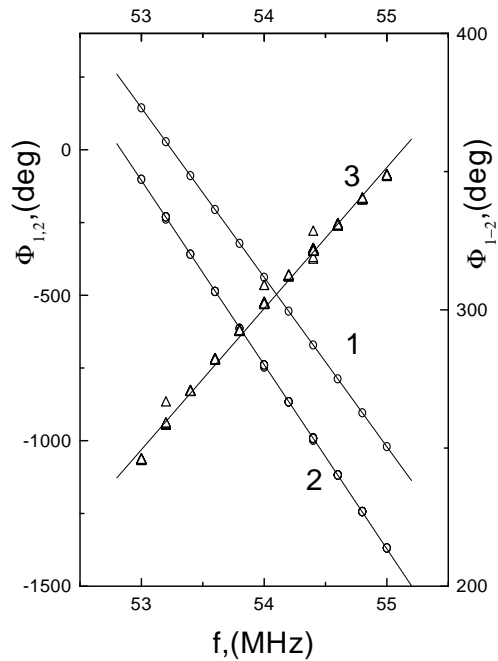}
\caption[]{Phase--frequency characteristics of the delay lines
({\em 1}), of a sandwich consisting of the sample (LuNi$_{2}$B$_{2}$C,
${\bf q}\parallel [100]$, ${\bf u}\parallel [100]$, $L = 0.835$ mm)
and the delay lines ({\em 2}), and the difference function, i.e.,
the P--F characteristic of the sample ({\em 3}). Note the difference
in the scales of the vertical axes. }
\label{f2}
\end{figure}

The measurement algorithm in part resembles one proposed
earlier.\cite{6} First the phase--frequency (P--F) characteristic
of an acoustic circuit consisting of two delay lines is measured
at fixed frequency points (step 1). Then the P--F characteristic
of a sandwich consisting of the same delay lines but with the sample
between them (Fig.\ \ref{f2}) is measured at the same temperature
(step 2).\footnote{At the frequencies we used the scale of the phase
variations of the signal are much greater than 360$^{\circ}$.
The phase meter, of course, measures phase differences in the interval
0--360$^{\circ}$, and the absence of discontinuities
(360$^{\circ}$ jumps) in Fig.\ \ref{f2} is achieved
through programming. }

Because the signal circuits contain elements capable of resonating
(piezotransducers, imperfectly matched feeders), each of these
characteristics is not necessarily a straight line. However, their
difference, i.e., the P--F characteristic of the sample, in the absence
of interference distortions in it, should form a strictly straight
line, the slope of which determines the phase velocity of the sound,
\begin{equation}
v=\frac{360 L}{S}
\end{equation}
where $v$ is the sound velocity (cm/s), $L$ is the thickness of the
sample (cm), and $S$ is the slope of the P--F difference characteristic
(deg/Hz). It is easily seen by a direct calculation that when the P--F
characteristics {\em 1} and {\em 2} are approximated by straight lines
by the least-squares method (the slopes are $S_{1}$ and $S_{2}$,
respectively), then
\begin{equation}
S=S_1 - S_2
\end{equation}
for any deviations of the P--F characteristics {\em 1} and {\em 2} from
straight lines. This relation is valid only if the frequency points at
which the P--F characteristics {\em 1} and {\em 2} are measured are
coincident. In Ref.\ \onlinecite{6} essentially the same procedure was
used to determine $S$, but since the technique used there did not
guarantee the required coincidence, additional errors
could have been introduced.

If $S$ is comparable to $S_{1}$ (0.3 or larger), then in homogeneous
materials the measurements can be limited to this step with completely
acceptable accuracy (0.3\% or better).

However, in homogeneous but rather thin samples the superposition of
secondary reflections distorts the main part of the measurement signal.
Because of this, the parts of the pulse that coincide with the leading
edge are customarily used for measurements. An analogous procedure, as
a rule, should be used in inhomogeneous materials for the reasons
already mentioned, even though the acoustic path length in them may be
comparatively large.

As a result of the occurrence of various kinds of transient processes,
the rate of which depends on the carrier frequency of the pulses, the
slopes of the P--F characteristics {\em 1} and {\em 2} become functions
of the temporal position of the strobe readout pulse at the leading
edge of the measurement signal. The variation of $S_{1,2}$, depending
on the type of piezotransducers, is 2--4\% (for comparison, in extended
samples the variation of $S_{1,2}$ at the steady part of the pulse is
at the 0.1\% level). This means that in going from step 1 to step 2 the
readout pulse should be shifted precisely by the sound delay time
$\tau _0$ in the sample. Since the latter is initially unknown and also
because of the discreteness of the step for the time shift of the strobe
signal ($5\times 10^{-8}$ s in our experiments), it was practically
impossible to satisfy this condition. For finding $\tau _0$ (and,
hence, the sound velocity) we used the following interpolation
procedure.

\begin{figure}
\includegraphics[width=7cm,angle=0]{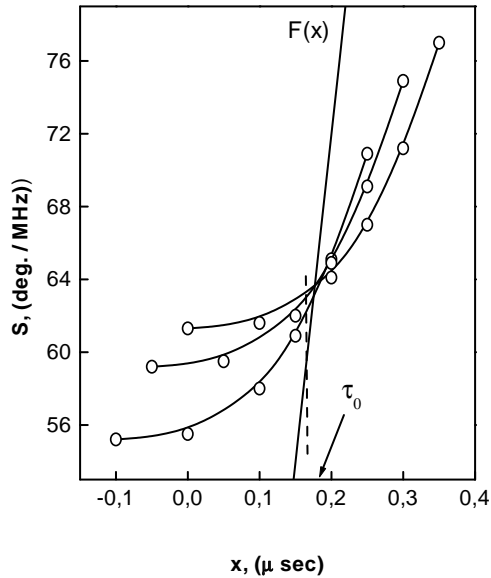}
\caption[]{Example of the interpolation procedure for finding the sound
delay time $\tau _0$. YNi$_{2}$B$_{2}$C sample (${\bf q}\parallel
[100]$, ${\bf u}\parallel [010]$, $L = 0.885$ mm) for several values of
$t_{c}$ (see text). At $x = 0$ the values of $t_{c}$ increase from
bottom to top with a step of $5\times 10^{-8}$ s. The linear function
$F(x) = 360x$. }
\label{f3}
\end{figure}

For each series of measurements with a definite mode (longitudinal or
transverse) we calibrated the dependence of $S_{1}$ on the temporal
position $t_{x}$ of the readout pulse. Then for a given sample we
measured $S_{2}$ at some known position $t_{c}$ of the readout pulse at
the leading edge of the signal. It is easy to see from Eqs.\ (1) and
(2) that $\tau _0$ is a solution of the equation $S(x) = 360x$, where
$x\equiv t_{c}-t_{x}$ is the time shift of the readout pulse between
the set of calibration measurements $S_{1}$ and the measurements
with the sample, $S_{2}$. An example of the graphical solution of the
interpolation equation for several values of $t_{c}$ is presented in
Fig.\ \ref{f3}. The results of the interpolation (the value of $\tau
_0)$ coincide regardless of the choice of $t_{c}$.

At this step of the procedure the ``rough'' determination of the sound
velocity is completed. To refine the values we use the ``Nonius''
method. Let the phase of the signal registered at some definite
frequency $f_0$ by the phase meter in step 1 be equal to $\Phi _{1}$.
In step 2 at the same frequency the phase of the signal will be $\Phi
_{2}$. The total phase inserted by the sample is $\Phi _0 = 360n+(\Phi
_{2}-\Phi _{1}$), where $n = 0,1,2,\ldots\,\,$. Since $\Phi _0 =
360f_0L/v$, by trying values of $n$ we find the refined value of $v$
that is closest to the ``rough'' estimate.

In the above discussion it was tacitly assumed that on going from step
1 to step 2 the phase of the signal changes only because of the
addition of the sample. Actually, however, besides the sample we also
had an additional layer of grease in step 2. During measurements in
very thin samples the contribution of the grease layer can become
noticeable. In our experiments GKZh-94 silicone oil was used as the
bonding agent, forming a layer ${\sim }1$--$2$ $\mu$m thick between
the ground surfaces. The passage of an elastic wave through such
a thin layer is described by the sum of an infinite geometric
progression with the denominator $q = k^{2}{\rm e}^{-2l(\alpha +iq)}$,
where $k$ is the reflection coefficient at the boundary (we assume that
the wave impedances of the delay line and sample are close in value),
$l$ is the thickness of the grease layer, $\alpha $ is the damping
coefficient, and $q$ is the wave number.

\begin{figure}
\includegraphics[width=7cm,angle=0]{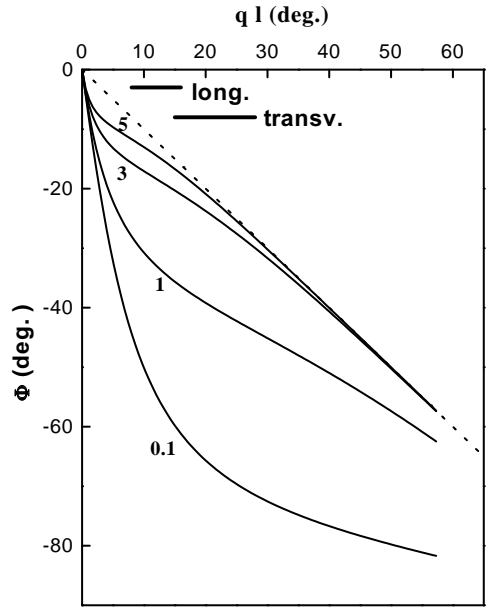}
\caption[]{Diagram pertaining to the calculation of the additional
phase shift inserted by the grease layer. The reflection coefficient at
the grease--sample boundary $k = 0.85$, and the numbers on the curves
give the sound damping coefficient in the grease layer (neper/cm). The
dotted curve corresponds to $\Phi  = -ql$. The horizontal lines
indicate the regions of the actual values of the parameter $ql$ for the
corresponding mode. }
\label{f4}
\end{figure}

An estimate of the propagation velocity of sound in the grease gave
$v_{l}\sim 2.1\times 10^{5}$ cm/s, $v_{t}\sim 1.2\times 10^{5}$ cm/s,
which correspond to reflection coefficients $k\sim 0.85$ for our
samples. In Fig.\ \ref{f4} we present the calculated dependence of the
phase of the wave passing through the grease layer on the thickness for
various damping coefficients. The regions of $ql$ corresponding to the
conditions of the experiment are also indicated in Fig.\ \ref{f4}.

At low damping the correction can be rather large. We were unable to
estimate the value of the sound damping in the grease --- in thick
layers (${\sim }0.5$ mm) it was very large, probably because of
cracking --- but we assume that its value is found at the 20
dB/cm level or higher, i.e., the phase inserted by the grease
layer is close to $ql$. In processing the results of the
measurements we introduced a correction for the additional grease
layer --- 10$^{\circ}$ for longitudinal sound and 20$^{\circ}$
for transverse sound. In thin samples the effect of this correction
was not over 1\%. We suppose
that this correction can be eliminated by making comparative
measurements on two samples of different thickness.\cite{3} In that
case the length difference $\delta L$ should be comparable to $L$,
since otherwise the contribution of possible nonuniformities of the
sound velocity over the whole length of the sample would be attributed
to the small difference $\delta L$.

Let us conclude with an estimate of the potential accuracy of a single
measurement. Special studies have established that the irreducibility
of the phase upon the remounting (regluing) of the acoustic
circuit is at the level of 20$^{\circ}$. We estimate the
indeterminacy of the correction for the additional grease layer to be
10$^{\circ}$. Assuming that the accuracy of the ``rough'' estimate of
the velocity is sufficient for determining the necessary value of $n$,
we obtain for the measurement error (at $f_0\sim 50$ MHz)
\begin{equation}
\frac {\delta v} {v} = \frac {30} {\Phi_0} \approx 2 \cdot 10^{-9}
\frac {v} {L}
\end{equation}

\section*{3. ELASTIC CONSTANTS OF BOROCARBIDES
${ \rm RN \lowercase{i}_{2}B_{2}C}$
(${\rm R = Y,Lu,  Ho} $)}

\begin{figure}
\includegraphics[width=8cm,angle=0]{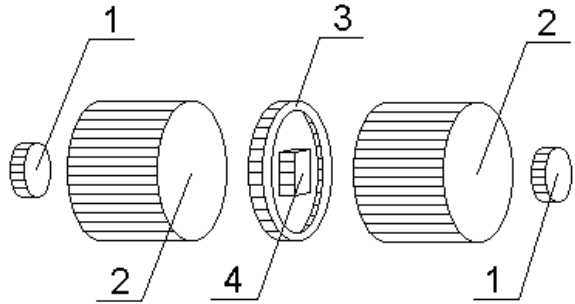}
\caption[]{Diagram of the mounting of the sample: 1 ---
piezotransducers, 2 --- delay lines, 3 --- brass support ring,
4 --- sample. }
\label{f5}
\end{figure}

In spite of the significant interest in the family of superconducting
borocarbides, very little information about their elastic properties
can be found in the literature. We know of only one ``acoustical''
study,\cite{7} devoted to YNi$_{2}$B$_{2}$C, in which the sound
velocity was measured by a time-of-flight method. Single crystals of
borocarbides were grown by the method described in Ref.\ \onlinecite{8}
and had the shape of a slab with a maximum dimension along the [001]
axis of ${\sim }0.8$ mm (${\rm R = Y}$), ${\sim }0.2$ mm (${\rm R =
Ho}$), and ${\sim }0.4$ mm (${\rm R = Lu}$). They were quite brittle,
and therefore the mounting of the samples between the delay lines was
done with the aid of a special brass ring, which acted as a holder and
reinforcer; the ring was ground simultaneously with the preparation of
the working faces (Fig.\ \ref{f5}). The diameter of the ring was chosen
larger than the diameter of the piezotransducers to prevent spurious
signals.

All of the measurements were made at liquid nitrogen temperature. The
results are presented in Table \ref{t1}. It contains some ``superfluous''
data, marked by an asterisk $(^{\ast })$. For example, for $C_{44}$ it
was sufficient to make a single measurement ${\bf q}\parallel [100]$,
${\bf u}\parallel [001]$ (${\bf u}$ is the polarization vector of the
elastic wave). We assume, however, that having the ``superfluous'' data
will make it possible to get an idea of the accuracy of the measurements
in this case.

One can also see that certain relations which follow from the general
theory of elasticity\cite{9} are well satisfied. For example, in a
tetragonal crystal the sum of the squares of the velocities of the
three modes remains constant under rotation of the wave vector ${\bf
q}$ in the (001) plane.

The elastic constants of the single crystals studied are presented in
Table \ref{t2}. The x-ray densities were used in calculating them. For
${\rm R = Y}$ the agreement with the results of Ref.\ \onlinecite{7} is
poor, although the relationships among the various constants are
preserved on the whole. The Debye temperature was calculated according
to the formula\cite{9}

\begin{equation}
\theta_{D} = 1146.8 \biggl( \frac {\rho s} {A I} \biggr) ^{1/3}
\end{equation}

where $A$ is the molecular weight, $s$ is the number of atoms in the
molecule, $\rho $ is the mass density, and $I$ is the sum of the inverse
cubes of the phase velocities of the elastic waves, averaged over all
directions of the wave normal. For ${\rm R = Ho}$, because of the
difficulty of preparing a sample of the required orientation, the
elastic constant $C_{13}$ was not measured, and in the calculation of
the bulk modulus and Debye temperature it was assumed equal to the
value of $C_{13}$ in lutecium borocarbide. For ${\rm R = Y}$ the
calculated value of $\theta _{D}$ is close to the thermodynamic
estimate.\cite{10} For ${\rm R = Lu}$ the deviation of the calculated
value of $\Theta _{D}$ from the thermodynamic value is, generally
speaking, greater than the allowable error. That may be an indication of
the existence in lutecium borocarbide of a low-temperature ferroelastic
structural transition, accompanied by a significant softening of some
elastic constant. Our preliminary measurements in holmium
borocarbide have shown that at 5.2 K the velocity of the $C_{66}$ mode
falls to ${\sim }3.3\times 10^{5}$ cm/s. When this softening is taken
into account, one obtains $\theta _{D} = 383$ K for ${\rm R = Ho}$.

This study was partly supported by the Government Foundation for Basic Research
of the Ministry of Education and Science of Ukraine (Grant No.\
0207/00359).

\begin{table}
\caption[]{Sound velocity in single crystals of
borocarbides ($T = 77$ K). }
\label{t1}
\vskip 1cm
\begin{tabular}{|c|c|c|c|c|}
\hline
\multicolumn{2}{|c|}{ Polarization} &
\multicolumn{1}{|c|}{YNi$_2$B$_2$C} &
\multicolumn{1}{|c|}{LuNi$_2$B$_2$C} &
\multicolumn{1}{|c|}{HoNi$_2$B$_2$C}
\\ \hline
$\bf {q \| }$ & $\bf {u \|}$ &  $v \cdot 10^5$cm/sec &
$v \cdot 10^5$cm/sec & $v \cdot 10^5$cm/sec
\\ \hline
[100] & [100] & 6.78 (0.885) & 5.88 (0.8) & 6.04 (0.606)
\\ \cline{2-5}
 & [001] & 3.25 (0.885) & 2.65 (0.8) & 2.73 (0.606)
\\ \cline{2-5}
 & [010] & 4.80 (0.885) & 4.30 (0.8) & 4.33 (0.606)
\\ \hline
[110] & [110]$^{\ast }$ & 7.55 (0.59) & 6.64 (0.988) & 6.86 (0.525)
\\ \cline{2-5}
 & [001]$^{\ast }$ & 3.26 (0.59) & 2.64 (0.988) & -
\\ \cline{2-5}
 & [$ 1 \overline{1} 0$] &  3.34 (0.59) & 2.77 (0.988) & 2.83 (0.525)
\\ \hline
[001] & [001] & 6.49 (0.84) & 6.01 (0.4) & 5.91 (0.23)
\\ \cline{2-5}
 & [100]$^{\ast}$ & 3.26 (0.84) & 2.70 (0.4) & 2.81 (0.23)
\\ \cline{2-5}
 & [010]$^{\ast }$ & 3.28 (0.84) & 2.70 (0.4) & 2.83 (0.23)
\\ \hline

45$^{\circ}$ from the [001] axis & QL$^{\ast}$ & 7.28 (0.303) & - & -
\\ \cline{2-5}
in the (110) plane & QT & 3.18 (0.465) & - & -
\\ \cline{2-5}
 & [110]$^{\ast}$ & 3.31 (0.303) & - & -
\\ \hline
45$^{\circ}$ from the [001] axis & & & &
\\
in the (100) plane & QT & - & 2.01 (0.27) & -
\\ \hline
\end{tabular}
\end{table}
\vskip 1cm
Note: The ``superfluous'' data are denoted by an asterisk ($^{\ast }$).
QL and QT are the quasilongitudinal and quasitransverse modes; the
thickness of the sample in mm is given in parentheses.

\begin{table}
\caption[]{Calculated parameters for borocarbides ($T = 77$ K) }
\vskip 1cm
\label{t2}
\begin{tabular}{|c|c|c|c|c|}
\hline
\multicolumn{1}{|c|}{Parameters } &
\multicolumn{2}{|c|}{YNi$_2$B$_2$C} &
\multicolumn{1}{|c|}{LuNi$_2$B$_2$C} &
\multicolumn{1}{|c|}{HoNi$_2$B$_2$C}
\\ \hline
C$_{11}$ &27.94  &22[7]  &29.39  &29.47
\\ \hline
C$_{12}$&14.39&9.84[7]&16.34&16.53
\\ \hline
C$_{13}$&17.81& &23.15&
\\ \hline
C$_{33}$&25.61&21.1[7]&30.68&28.20
\\ \hline
C$_{44}$&6.43&5.42[7]&5.97&6.02
\\ \hline
C$_{66}$&14.00&13.1[7]&15.71&15.15
\\ \hline
B&20.16& &20.27&23
\\ \hline
$\theta_{D}$, K&501&490[10]&409 (360[10])&404
\\ \hline
$\rho$, g/cm$^3$&6.08&6.05[7]&8.5&8.08
\\ \hline
\end{tabular}
\end{table}
\vskip 1cm
Note: $C_{ik}$ are elastic constants (in units of $10^{11}$
dyn/cm$^{2}$), $\theta _{D}$ is the Debye temperature, and $B$ is the bulk
modulus. For Ho the modulus C$_{13}$ was not measured, and in the
calculation of $\theta _{D}$ and $B$ it was assumed equal to 23.15 (see
text).

\end{document}